\documentstyle[12pt,epsfig]{article}
\begin{document}

\begin{titlepage}
\rightline{November 2012}
\vskip 2cm
\centerline{\Large \bf
Differentiating hidden sector dark matter 
}
\vskip 0.5cm
\centerline{\Large \bf 
from light WIMPs with Germanium detectors}

\vskip 2.2cm
\centerline{\bf R. Foot\footnote{ E-mail address: rfoot@unimelb.edu.au}}

\vskip 0.7cm
\centerline{\it ARC Centre of Excellence for Particle Physics at the Terascale,}
\centerline{\it School of Physics, University of Melbourne,}
\centerline{\it Victoria 3010 Australia}
\vskip 2cm
\noindent
Light WIMP dark matter and hidden sector dark matter have been proposed to explain the
DAMA, CoGeNT and CRESST-II data.  Both of these approaches feature spin independent 
elastic scattering of dark matter particles on nuclei.  Light WIMP 
dark matter invokes a single particle species which interacts with ordinary matter via 
contact interactions.  By contrast hidden sector dark matter is typically 
multi-component and is assumed to interact via the exchange of a massless mediator.
Such hidden sector dark matter thereby predicts a sharply rising nuclear recoil spectrum,
$dR/dE_R \sim 1/E_R^2$ due to this dynamics, while WIMP dark matter predicts a spectrum which
depends sensitively on the WIMP mass, $m_\chi$. 
We compare and contrast these two very different possible origins of the CoGeNT low energy excess.
In the relevant energy range, the recoil spectra predicted by these two theories
approximately agree provided $m_\chi \simeq 8.5$ GeV - close to the value favoured from fits to the
CoGeNT and CDMS low energy data.
Forthcoming experiments including C-4, CDEX, and the MAJORANA demonstrator, 
are expected to provide reasonably precise measurements
of the low energy Germanium recoil spectrum, including the annual modulation amplitude,
which should differentiate between these two theoretical possibilities.

\end{titlepage}


\section{Introduction}

The search for dark matter via direct detection experiments continues to yield very exciting
positive results. DAMA\cite{dama1,dama2}, CoGeNT\cite{cogent,cogent2} and CRESST-II\cite{cresst-II} 
have all reported results consistent with dark matter interactions. 
Of these, the DAMA annual modulation signal is (currently) the most convincing
evidence for dark matter direct detection. With more than 12 annual cycles of data collected, the annual modulation
amplitude deviates from zero at more than $8\sigma$ C.L. with both the phase and
period consistent with dark matter expectations\cite{dm}. 
These are clearly very interesting times for dark matter direct detection.

Low threshold experiments  
can probe specific dark matter explanations of the DAMA annual modulation signal. 
Initial results of the CoGeNT experiment have provided some tantalizing results.
Future data from 
C-4\cite{c4}, CDEX\cite{cdex} and the MAJORANA demonstrator\cite{maj}, all using a Germanium target, should
be able to confirm DAMA's direct detection. 
Furthermore, these experiments 
have the potential to measure both the recoil spectrum (i.e. unmodulated part) and annual modulation signal with high precision and
thereby distinguish between possible dark matter candidates.

Two quite different dark matter schemes have been proposed to explain the sharply rising 
low energy excess seen by CoGeNT.  One possibility is that this rising
event rate is due to interactions of a WIMP with contact (point-like) interaction\cite{review}.
Although such an interaction predicts a flat nuclear recoil energy spectrum at low energies:
$dR/dE_R \propto$ constant as $E_R \to 0$,
CoGeNT's low energy excess can be due to
kinematic effects. For a given nuclear recoil energy $E_R$, only a portion of the dark 
matter particles in the halo
have sufficient energy to produce the recoil. 
As $E_R$ decreases this portion exponentially increases, yielding a sharply 
rising event rate at low energies.
For low enough $E_R$, the rate should eventually flatten out producing $dR/dE_R \propto$ constant, assuming this dark matter
model is correct.
Since the recoil energy dependence of the spectrum is due to kinematic effects, CoGeNT
can sensitively probe the WIMP mass in this model, which comes out to around $10$ GeV\cite{c1}. 
While light WIMPs can also potentially explain DAMA\cite{gondolo} and CRESST-II data\cite{cresst-II},
it appears to be a major challenge to explain all three experiments simultaneously for a consistent 
set of parameters\cite{wimp}.

There is another explanation for the sharply rising excess seen by CoGeNT. 
This excess might be
due to dark matter particles interacting with ordinary matter via a massless 
mediator\cite{foot69}.  Dark matter arising from a hidden sector, with $U(1)'$ gauge interaction, 
coupling with the ordinary matter via
$U(1)'$ - $U(1)_Y$ kinetic mixing is one possibility\cite{he,flv}.
If the $U(1)'$ is unbroken then the mediator is massless, and can be recognized as the photon.
In this type of dark matter theory, the cross-section has a non-trivial recoil
energy dependence, $d\sigma/dE_R \propto 1/E_R^2$.
This implies $dR/dE_R \propto 1/E_R^2$ at low energies which, it turns out\cite{foot2012a,foot2012b}, 
is compatible with CoGeNT's data.
In this case the dark matter particles can be heavier ($\stackrel{>}{\sim} 20$ GeV) as no kinematic effect
is required to explain the excess.  It further turns out that in this picture, denoted henceforth as
hidden sector dark matter, the DAMA, CoGeNT and CRESST-II data can be simultaneously
explained\cite{foot2012a,foot2012b}.

The purpose of this 
article is to compare and contrast these two quite distinct possible origins of the CoGeNT low energy excess.
In particular we focus on future Germanium experiments because, as discussed above, 
these have the potential to measure the low energy spectrum and annual modulation amplitude 
with high precision.

\section{Light WIMPs versus hidden sector dark matter}

\vskip 0.3cm
\noindent
{\it Light WIMPs}
\vskip 0.4cm

Dark matter consisting of WIMP
particles $\chi$, interacting with ordinary matter via a contact interaction is a simple and certainly popular
dark matter candidate.
The cross-section for such a WIMP, with velocity $v$, interacting with a Germanium target is\cite{review}
\footnote{Natural units, $\hbar = c = 1$, are used throughout.}
\begin{eqnarray}
{d\sigma \over dE_R} = {m_{Ge} \over 2v^2} {\sigma_n \over \mu^2_{n}}A^2 F^2 (q)
\label{obama}
\end{eqnarray}
where $\mu_n$ is the $\chi$-neutron reduced mass, $\sigma_n$ is the $\chi$-neutron cross-section
and $F(q)$ is the form factor. Also, $A$ is the mass number of the target nuclei, and isospin invariant $\chi$ interactions
have been assumed.

In addition to the cross-section, the
interaction rate in a direct detection experiment also depends on the dark matter galactic halo distribution.
Dark matter in the halo is generally assumed to be in a Maxwellian distribution:
\begin{eqnarray}
f_\chi ({\textbf{v}}_E, {\textbf{v}}) = e^{-E/T} = e^{-({\textbf{v}}_E + {\textbf{v}})^2/v_0^2}
\end{eqnarray}
where ${\textbf{v}}_E$ is the  
velocity of the Earth with respect to the halo, and ${\textbf{v}}$ is the velocity of the dark matter particles
with respect to the Earth. 
In the standard halo model, the effective temperature of the Maxwellian distribution scales as the square
of the galactic rotational velocity: $T = {1\over 2} m_\chi v_{rot}^2$.
Evidently $v_0 \equiv \sqrt{2T/m_\chi} = v_{rot}$.
We take the reference value
$v_{rot} = 230$ km/s since later we wish to compare with some results of ref.\cite{cf}.
Also, $|{\textbf{v}}_E| = v_{\odot} + v_{orb} \cos\gamma \cos \omega (t - t_0)$,
with $v_\odot = v_{rot} + 12$ km/s, $v_{orb} = 30$ km/s, $\cos\gamma = 0.5$ and $t_0 = 152.5$ days (June $2^{nd}$).
The velocity distribution is limited in this model by the 
galactic escape velocity, which we take as $600$ km/s.
That is, $|{\textbf{v}} + {\textbf{v}}_E| < 600$ km/s.

The rate for $\chi$ scattering on a Germanium target
nucleus is 
\begin{eqnarray}
{dR \over dE_R} = N_T n_{\chi} 
\int_{|{\textbf{v}}| > v_{min}}
{d\sigma \over dE_R}
{f_{\chi}({\textbf{v}},{\textbf{v}}_E) 
\over k} |{\textbf{v}}| d^3 {\textbf{v}} 
\label{55}
\end{eqnarray}
where the integration limit is
$\ v_{min} \ = \ \sqrt{ (m_{Ge} + m_{\chi})^2 E_R/2 m_{Ge} m^2_{\chi} }$\ .
In Eq.(\ref{55}), $k=v_0^3 \ \pi^{3/2}$, $N_T$ is the number of target nuclei and 
$n_{\chi} = \rho_{dm} \xi_{\chi}/m_{\chi}$ 
is the number density of the halo $\chi$ particles. 
[$\rho_{dm} = 0.3 \  {\rm GeV/cm}^3$ and $\xi_{\chi}$ is the halo mass fraction
of species $\chi$, generally assumed to be unity in single component dark 
matter models].

\vskip 0.4cm
\noindent
{\it Hidden sector dark matter}
\vskip 0.4cm

The other class of models we consider is where dark matter interacts with ordinary matter
via a massless mediator. 
Dark matter from a hidden sector
with $U(1)'$ gauge interaction, coupling with ordinary matter via $U(1)' - U(1)_Y$ kinetic
mixing\cite{he,flv} is a simple renormalizable example of such a model (which we henceforth adopt).
If the $U(1)'$ is unbroken then the kinetic mixing induces a tiny ordinary electric charge for the $U(1)'$ charged
particles\cite{holdom}. This induced charge, denoted by $\epsilon e$,
enables the hidden sector particles to elastically scatter off ordinary charged particles
such as the nuclei in atoms. The cross-section for such a hidden sector  particle of velocity $v$ to thereby elastically 
scatter off a Germanium nucleus is given by\cite{foot69}:
\begin{eqnarray}
{d\sigma \over dE_R} = {2\pi \epsilon^2 Z_{Ge}^2 \alpha^2 F^2 (q) \over m_{Ge} E_R^2 v^2}
\label{cs}
\end{eqnarray}
where $Z_{Ge} = 32$ is the atomic number of Germanium and $\alpha$ is the fine structure constant.
The most important difference between
this Rutherford-type cross-section, and the one for standard WIMPs, Eq.(\ref{obama}), is the $1/E_R^2$ dependence.
It arises because the Feynman diagram for the elastic scattering process involving the exchange of a massless
mediator\footnote{If the mediator is not massless, but has mass $m$, then the Feynman amplitude is proportional
to $1/(q^2 + m^2)$. Thus the cross-section is point-like or Rutherford-like depending on 
whether $q^2 \ll m^2$ or $q^2 \gg m^2$. For the relevant recoil energies, the cross-section is
approximately Rutherford-like, $d\sigma/dE_R \propto 1/E_R^2$,
provided $m \stackrel{<}{\sim} 10$ MeV. See ref.\cite{ital} for a study of the effect of varying $m$.}
has amplitude proportional to $1/q^2 \simeq 1/(2m_{Ge}E_R)$.  

Hidden sector dark matter has a number of other very distinctive properties. 
Firstly, such dark matter
can have significant self-interactions mediated by the unbroken $U(1)'$ gauge interaction.
Secondly, it is also dissipative since a plasma composed of such particles can lose energy via radiating the $U(1)'$
`dark' photon in bremsstrahlung processes.
A third distinctive feature of these kinds of models is that they
are necessarily multi-component if the dark matter in the Universe arises from 
a particle-antiparticle asymmetry. 
This is due to the $U(1)'$ neutrality of the Universe.
See ref.\cite{foot2012b}
for further details and also ref.\cite{feng,kali} and references there-in for some 
relevant astrophysical/cosmological discussions of closely related
models.

We consider the simplest such multi-component hidden sector model with two stable 
hidden sector particles, $F_1$ and $F_2$\ \footnote{
It is supposed that the binding energy of any bound states that $F_1$ and $F_2$ might
form is much less than the halo temperature $T$, a situation
that is easy to satisfy. With this condition the halo is composed predominately of 
unbound $F_1$ and $F_2$ particles. 
The alternative case, where $F_1$ and $F_2$ form tightly bound `dark atoms' has
quite different phenomenology\cite{cline}.}. 
The hydrostatic equilibrium condition on an isothermal spherical distribution of such particles
implies\cite{foot69,sph}:
\begin{eqnarray}
v_0[F_i] = \sqrt{{2T \over m_{F_i}}} 
        = v_{rot} \sqrt{{\bar m \over m_{F_i}}}\  
\label{v0}
\end{eqnarray} 
where 
$\bar m$ is the mean mass of the
particles in the halo.
Thus for hidden sector dark matter, their halo distribution is still expected
to be Maxwellian but $v_0 \neq v_{rot}$. 
If $m_{F_2} \gg m_{F_1}$ it is possible to have $m_{F_2} \gg \bar m$,
and hence $v_0 [F_2] \ll v_{rot}$.
This narrow velocity dispersion of $F_2$ can lead to a situation where
lower threshold experiments, such as DAMA, CoGeNT are able to see 
$F_2$ interactions while higher threshold experiments such as XENON100\cite{xenon100} and CDMS\cite{cdms} do not.
This can help explain why XENON100 and CDMS have yet to find a positive signal, even with
$F_2$ as heavy as $\sim 50$ GeV\cite{foot2012b}.

The scattering rate of hidden sector $F_2$ particles on target nuclei is of the same form
as in Eq.(\ref{55}),  but with cross-section given in Eq.(\ref{cs}) and the $v_0$ 
value given in Eq.(\ref{v0}).  Note that for hidden sector dark matter, with 
significant self-interactions, the velocity distribution
is not constrained by a galactic escape velocity limit.  
\vskip 0.4cm
\noindent
{\it Mirror dark matter}
\vskip 0.4cm

Mirror dark matter corresponds to the interesting special case where the hidden sector is isomorphic to 
the standard model sector. 
The hidden sector, which we refer to as the mirror sector in this case, thus has gauge
symmetry $SU(3)'\otimes SU(2)'_L \otimes U(1)'_Y$. 
The Lagrangian describing the ordinary and hidden sectors also respects 
an exact and unbroken $Z_2$ mirror symmetry which can be interpreted as space-time parity\cite{flv}. 
The $Z_2$ mirror symmetry interchanges each ordinary particle (scalar, fermions and gauge bosons) with a corresponding partner,
denoted with a prime ($'$).
This discrete symmetry ensures that the fundamental properties of the hidden
sector particles exactly mirror those of the ordinary sector. 
There is thus a spectrum of `mirror' particles
e$'$, H$'$, He$'$, O$'$, Fe$'$, .... The $Z_2$ mirror symmetry ensures that 
the mass of each mirror particle is the same as the corresponding ordinary particle.
For a more extensive treatment, including astrophysical and cosmological discussions, see the reviews\cite{review2} 
and references there-in.

Since the mirror sector is isomorphic to the standard model sector, the theory
contains two gauged $U(1)$ symmetries, $U(1)_Y$ and $U(1)'_Y$.
As discussed above for the generic hidden sector case, these two $U(1)'s$ can kinetically mix
thereby inducing a small ordinary electric charge for the hidden sector
charged particles.
Thus, a mirror nucleus, $A'$, with atomic number $Z'$  will couple to ordinary photons with 
electric charge $\epsilon Z'e$.
Such a mirror nucleus moving with velocity $v$
can thereby elastically scatter off an ordinary nucleus, $A$, with
atomic number $Z$. This imparts an observable recoil energy, $E_R$, with
\begin{eqnarray}
{d\sigma \over dE_R} = {2\pi \epsilon^2 Z^2 Z'^2 \alpha^2 F^2_A F^2_{A'} \over m_A E_R^2 v^2}
\label{cs2}
\end{eqnarray}
where $F_A$ [$F_{A'}$] is the form factor 
of the nucleus [mirror nucleus]. 

In this theory, the galactic dark matter halo of the Milky Way is composed predominately of mirror particles. 
These particles form a pressure supported, multi-component plasma
containing e$'$, H$'$, He$'$, O$'$, Fe$'$,...\cite{sph}.  
Each particle species is
described by a Maxwellian distribution, but with velocity dispersion depending
on the mass of the component,  as in Eq.(\ref{v0}).
In the mirror dark matter case, $\bar m$ is not a free parameter, but constrained to be approximately $1.1 $ GeV
from mirror BBN calculations\cite{paolo2}.

\section{Germanium spectrum - unmodulated part}

Previous work\cite{foot2012a,foot2012b} has shown that the DAMA, CoGeNT and CRESST-II experiments can be simultaneously explained within
the generic hidden sector framework as well as the more specific mirror dark matter case.
These explanations are also consistent with the null results of other experiments
such as XENON100\cite{xenon100} and CDMS\cite{cdms}, although not without some tension. These null results do
suggest a narrow velocity dispersion of the detected dark matter particle $F_2$, i.e. $\bar m \ll m_{F_2}$ in the
two component hidden sector model considered here.
The mean mass $\bar m$ can be defined in terms of the abundances
of $F_1$ and $F_2$:
via $\bar m = (n_{F_1} m_{F_1} + n_{F_2} m_{F_2})/(n_{F_1} + n_{F_2})$. 
The $U(1)'$ charge neutrality of a plasma of such particles implies 
$q'_{F_1} n_{F_1} + q'_{F_2} n_{F_2} = 0$ where $q'_{F_j}$ is the $U(1)'$ charge of $F_j$ ($j=1,2$).
Thus the condition $\bar m \ll m_{F_2}$ implies that $m_{F_1} \ll m_{F_2}$ and $|q'_{F_1}| \ll |q'_{F_2}|$.
Such a situation is of course possible.
[As briefly mentioned in the previous section, mirror dark matter 
for instance, predicts $\bar m \approx 1.1$ GeV from mirror BBN calculations\cite{paolo2}].
The $m_{F_1} \ll m_{F_2}$ requirement suggests that the experiments are only directly sensitive to 
interactions of the $F_2$ component.

Fits to the DAMA annual modulation data implicate hidden sector dark matter mass in the range: 
$m_{F_2} \stackrel{>}{\sim} 20$ GeV (for $v_{rot} = 230$ km/s) \cite{foot2012b}. 
A rough upper limit, $m_{F_2} \stackrel{<}{\sim} 60$ GeV is suggested by the null
results of the XENON100 and CDMS direct detection experiments\cite{foot2012b}.
In figure 1 we examine the predicted Germanium recoil spectrum for $F_2$ 
masses in this range, with the kinetic mixing parameter $\epsilon$ 
adjusted so that the low energy normalization is fixed.

\vskip 0.3cm
\centerline{\epsfig{file=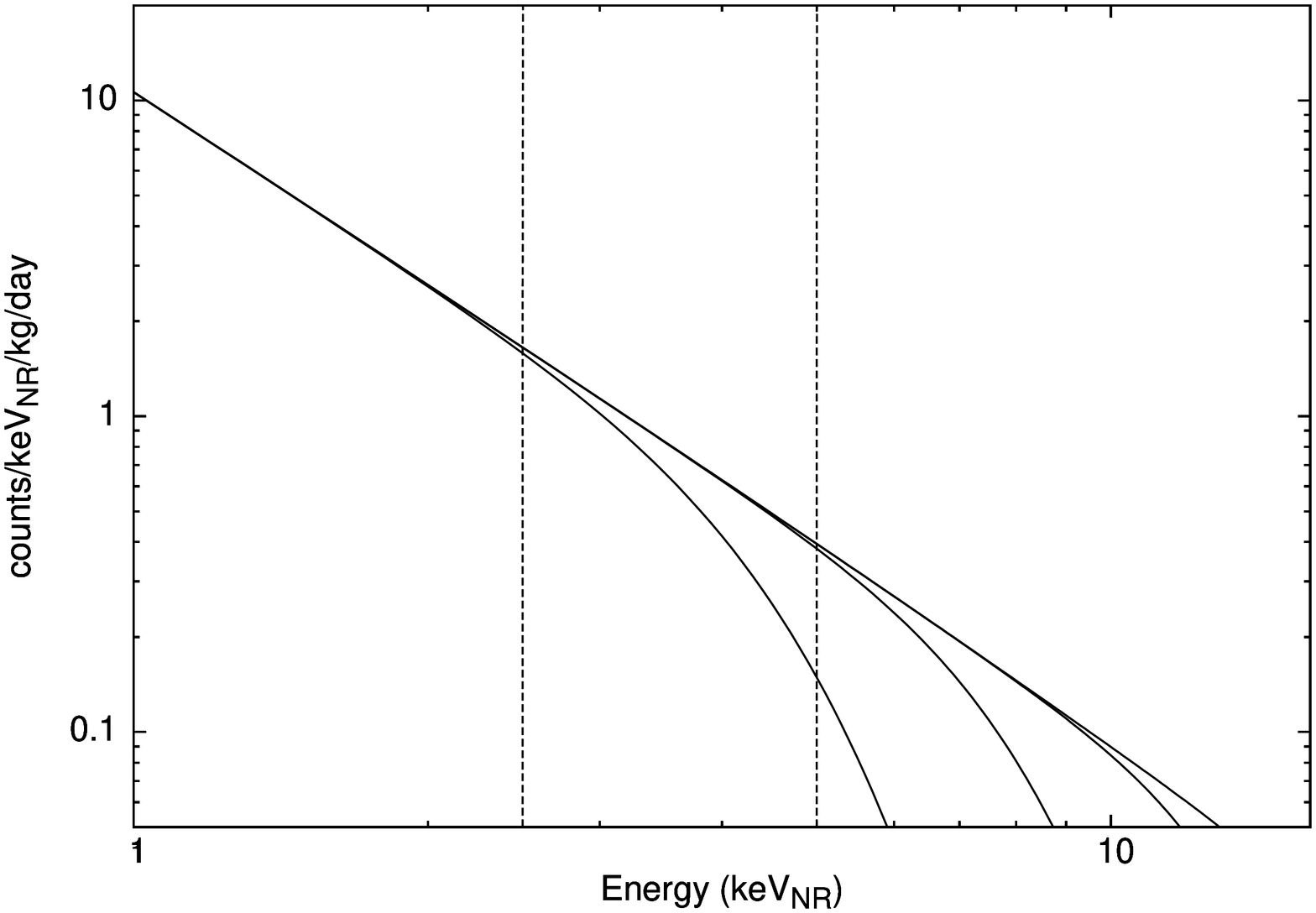,angle=270,width=12.9cm}}
\vskip 0.2cm
\noindent
{\small
Figure 1: Germanium recoil spectrum from hidden sector dark matter.
The lines from left to right correspond to $m_{F_2}/{\rm GeV}$ = 20, 30, 45, 60.
[$v_{rot} = 230$ km/s and $\bar m = 1.0$ GeV].
The vertical dashed lines show the range of CoGeNT's low energy excess.
}

\vskip 1.0cm

Figure 1 clearly demonstrates the expected $dR/dE_R \propto 1/E_R^2$ behaviour 
at low energies of hidden sector dark matter.
The energy threshold of the CoGeNT experiment is $0.5 \ {\rm keVee} \approx 2.5 \ {\rm keV}_{NR}$\footnote{
The keVee unit refers to the measured ionization energy while kev$_{NR}$
is the nuclear recoil energy equivalent.
We assume that the quenching factor, $q \equiv {\rm keV}_{NR}/{\rm keVee}$, is given by $q = 0.18 (E_R/{\rm keV}_{NR})^{0.12}$. 
This is in between the value suggested by  CoGeNT\cite{cogent2} and by the recent study\cite{bmei}. It is also consistent 
with the experimental measurements summarized in figure 5 of ref.\cite{cogent2}. The uncertainty in the quenching
factor, and thus CoGeNT's nuclear recoil energy scale, is around 10-20\%.}.
The CoGeNT low energy excess can be 
extracted from the background over the energy range:
$2.5 \ {\rm keV}_{NR} \stackrel{<}{\sim} E_R \stackrel{<}{\sim} 5 \ {\rm keV}_{NR}$.  
For $m_{F_2} \stackrel{>}{\sim} 30$ GeV, the $dR/dE_R \propto 1/E_R^2$ dependence extends throughout the entire 
CoGeNT low energy `signal' region. 
For $20 \ {\rm GeV} \stackrel{<}{\sim} m_{F_2} \stackrel{<}{\sim} 30 \ {\rm GeV}$ 
the shape of the spectrum can be somewhat steeper.
A modest preference for $m_{F_2} \stackrel{>}{\sim} 30$ GeV arises\cite{foot2012b} in this model from the relative
normalizations of the DAMA and CoGeNT signals.
With this justification,
we now focus attention on hidden sector dark matter with $m_{F_2} 
\stackrel{>}{\sim} 30$ GeV for which the recoil energy spectral shape is `predicted'.
[We comment later on the slight changes to the spectral shape if $m_{F_2}$ is in the lower mass window
$20 \ {\rm GeV} \stackrel{<}{\sim} m_{F_2} \stackrel{<}{\sim} 30 \ {\rm GeV}$].

\vskip 0.6cm
\centerline{\epsfig{file=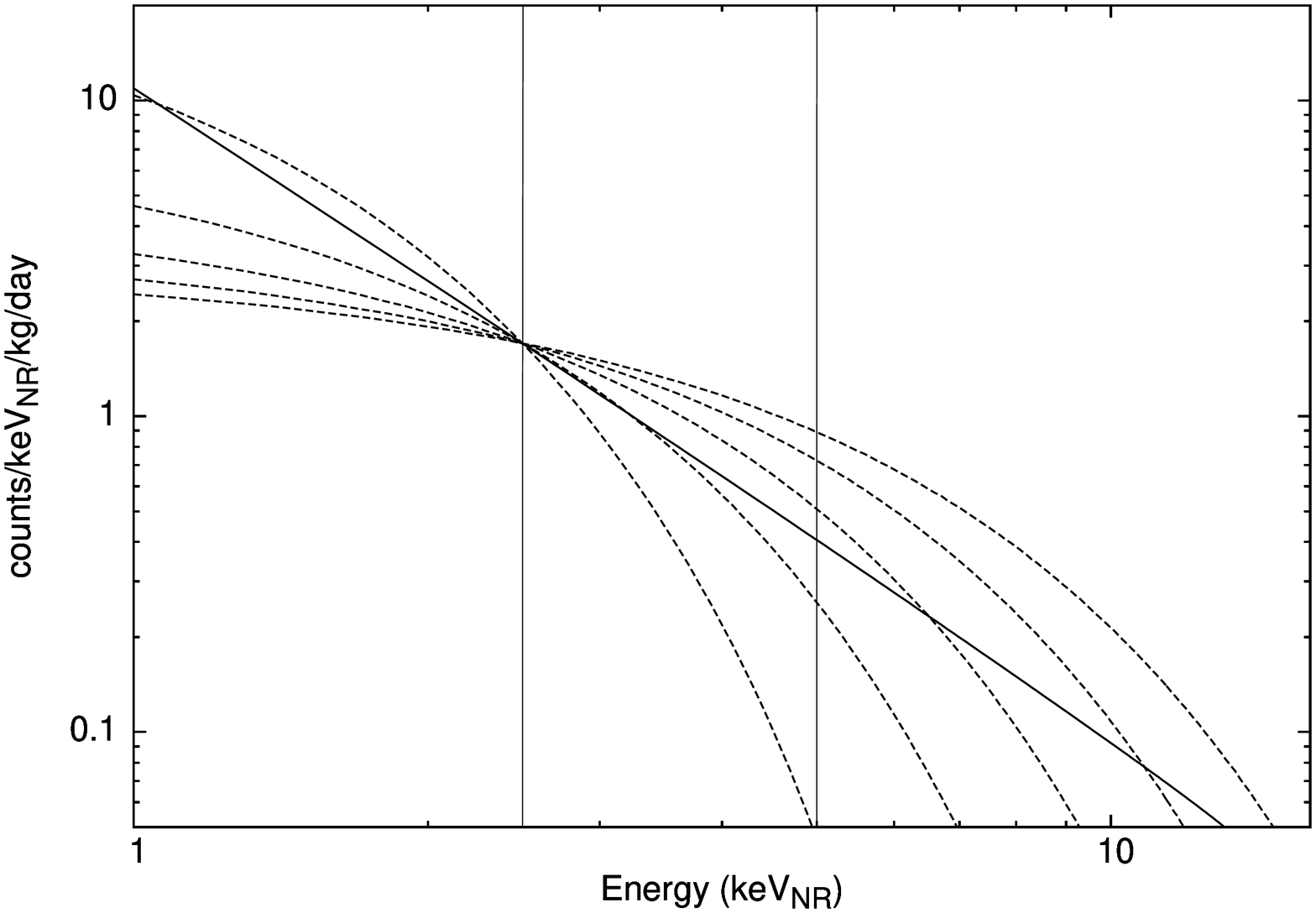,angle=270,width=12.9cm}}
\vskip 0.3cm
\noindent
{\small
Figure 2: Comparison of the Germanium recoil spectrum predicted by hidden sector dark matter 
with that from light WIMPs.
The solid line is the hidden sector prediction (for $m_{F_2} \stackrel{>}{\sim} 30$ GeV)
while dashed lines (from steepest to flattest) are predictions of light
WIMPs for $m_\chi/{\rm GeV} =$ 6, 8, 10, 12, 14.
The vertical dashed lines show the energy range of the CoGeNT excess and
$v_{rot} = 230$ km/s is assumed.
}

\vskip 1.4cm

In figure 2, we compare the Germanium recoil spectrum of hidden sector dark matter with
that obtained from light WIMPs.
We adjust  $\sigma_n$ so that the rates have fixed normalization at 2.5 ${\rm keV}_{NR}$.
Figure 2 shows that WIMPs with mass around $m_\chi \approx 8$ GeV produce a Germanium recoil spectrum of similar shape
to the $dR/dE_R \propto 1/E_R^2$ dependence characteristic of hidden sector dark matter.  
To quantify this, we can define the `area', ${\cal A} (m_\chi)$ as:
\begin{eqnarray}
{\cal A}(m_\chi) = \int^{5 \ {\rm keV}_{NR}}_{2.5 \ {\rm keV}_{NR}} \ \left| {dR^{lw} \over dE_R} - {dR^{hs} \over dE_R} \right| \ dE_R
\end{eqnarray}
where $dR^{lw}/dE_R$ 
[$dR^{hs}/dE_R$] 
is the rate in the light WIMP model [hidden sector model].
We find that ${\cal A}(m_\chi)$ is minimized when $m_\chi \simeq 8.5$ GeV. 
This means that for the energy range of CoGeNT's excess, the scattering of $m_\chi \simeq 8.5$ GeV WIMPs 
produces a recoil spectrum that most closely resembles
the $dR/dE_R \sim 1/E_R^2$ spectrum characteristic of hidden sector dark matter.
This is illustrated in figure 3, along with CoGeNT's data (corrected for efficiency, stripped of background components, and with
surface event correction\cite{cogent2}).

Repeating this analysis for the low $m_{F_2}$ mass window, 
$20 \ {\rm GeV} \stackrel{<}{\sim} m_{F_2} \stackrel{<}{\sim} 30 \ {\rm GeV}$
we find that ${\cal A}(m_\chi)$ is minimized
for $7.3 \ {\rm GeV}  \stackrel{<}{\sim} m_\chi \stackrel{<}{\sim} 8.5 \ {\rm GeV}$.
Thus, there is only a very narrow $m_\chi$ window where WIMPs can mimic the hidden sector dark matter
spectral shape for the relevant CoGeNT energies.

\vskip 0.2cm
\centerline{\epsfig{file=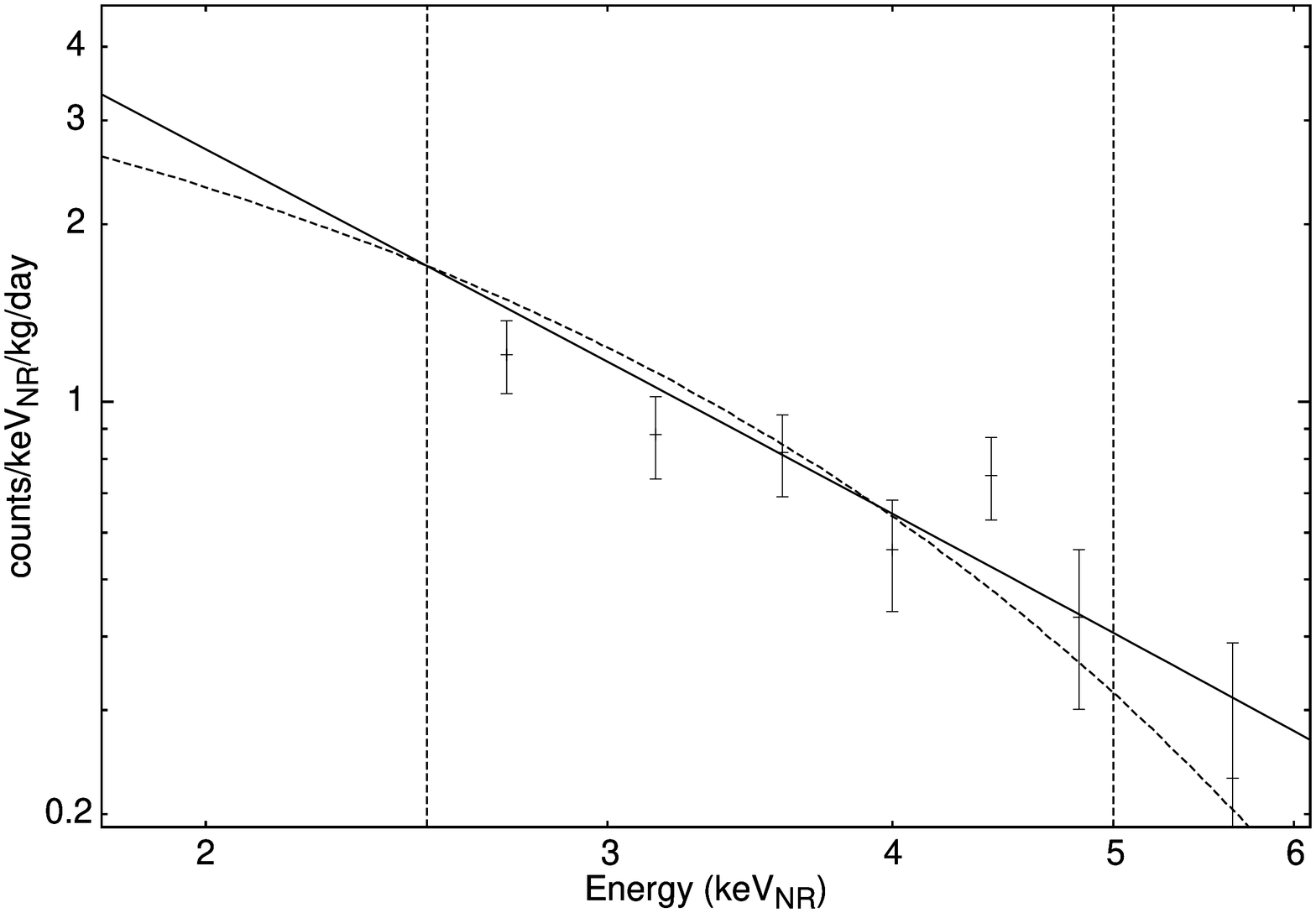,angle=270,width=12.6cm}}
\vskip 0.2cm
\noindent
{\small
Figure 3: Comparison of the Germanium recoil spectrum predicted by hidden sector dark matter 
(solid line)
with that from WIMPs with mass 8.5 GeV (dashed line). 
The vertical dashed lines show the energy range of the CoGeNT excess and
$v_{rot} = 230$ km/s is assumed. CoGeNT data, obtained from ref.\cite{cogent2}, is also shown.
}

\vskip 1.0cm

If hidden sector dark matter is correct then 
an analysis of the data in terms of light WIMPs
should be consistent with $7.3 \ {\rm GeV} \stackrel{<}{\sim} m_\chi \stackrel{<}{\sim} 8.5$ GeV.
Systematic uncertainties due to calibration issues
should be of order a GeV or less.
In figure 4, we show an analysis of the CoGeNT and CDMS data in the light WIMP model.
The CoGeNT analysis uses the most recent data from ref.\cite{cogent2} corresponding to 0.33 kg $\times$ 807 days. 
[This data is corrected for efficiency, stripped of known background components and
with surface event correction]. Quenching factor uncertainties are taken into account as in ref.\cite{foot2012b}.
The CDMS data refers to the
CDMS low energy data given in ref.\cite{cdmslow}. This data  covers roughly the same recoil
energy range as CoGeNT's signal region [$2.5 \ {\rm keV}_{NR} \stackrel{<}{\sim} E_R \stackrel{<}{\sim} 5 \ {\rm keV}_{NR}$]. 
Collar and Fields\cite{cf} carefully analysed this data and 
identified a family of events in the nuclear recoil band - a tentative
dark matter signal. The analysis of ref.\cite{cf} reveals that 
the CDMS experiment pins down $m_\chi$ fairly precisely due to its large
exposure.    
The analysis of CDMS low energy data of ref.\cite{cf} could be repeated in the hidden sector framework. In lieu of that, it is
useful to know the value of $m_\chi$ for which the spectrum of light WIMPs approximately mimics
that of hidden sector dark matter.
Figure 4 shows that both the CoGeNT and CDMS data are consistent with $m_\chi \simeq 8.5$ GeV.
This can be viewed as interesting evidence in support of hidden sector dark matter.

\vskip 0.3cm
\centerline{\epsfig{file=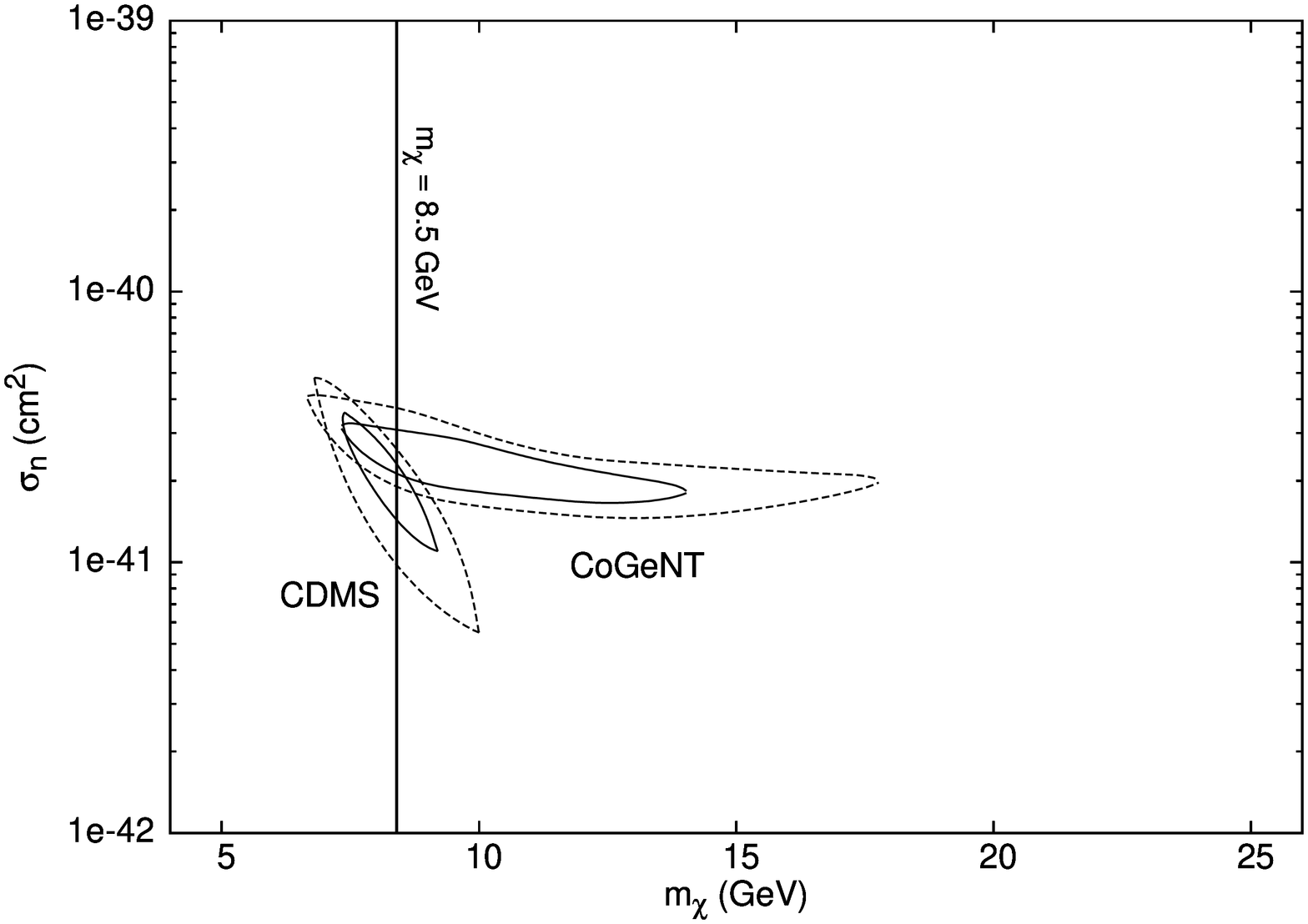,angle=270,width=12.9cm}}
\vskip 0.3cm
\noindent
{\small
Figure 4: Favoured $\sigma_n, \ m_\chi$ parameter region [$90\%$ and $99\%$ C.L.] in the light WIMP model for $v_{rot} = 230$ km/s. 
The CDMS allowed regions are from ref.\cite{cf}.
The vertical solid line indicates $m_\chi = 8.5$ GeV - identified as the WIMP mass where
the Ge recoil spectra most closely matches that of hidden sector dark matter ($m_{F_2} \stackrel{>}{\sim} 30$ GeV). 
}

\vskip 1.7cm


%

The discussion so far has fixed $v_{rot} = 230$ km/s.
Varying $v_{rot}$ away from this `reference' value will not change the shape predicted
in the hidden sector model. This is because the shape is governed by the $d\sigma/dE_R \propto 1/E_R^2$ dynamics.
However, for a fixed $m_\chi$, changing $v_{rot}$ will change the predicted
spectral shape in the light WIMP model. Thus, the value of $m_\chi$ where light WIMPs most closely mimic
hidden sector dark matter
will change depending on the value of $v_{rot}$ chosen (table 1).
However, the value of $m_\chi$ favoured by the data also changes and in roughly the same way. Thus our conclusions
remain unchanged if we vary $v_{rot}$.


\begin{table}
\centering
\begin{tabular}{c c c}
\hline\hline
$v_{rot}$  & $m_\chi$ & ${\cal A}_{min}\ {\rm kg}^{-1} {\rm day}^{-1}$  \\
\hline
200\ km/s  & 10.0 GeV  & 0.11  \\
230\ km/s  & 8.5 GeV  & 0.11  \\
260\ km/s  & 7.5 GeV  & 0.11  \\
\hline\hline
\end{tabular}
\caption{
The $m_\chi$ value for which the scattering of WIMPs produces a Ge recoil spectrum that 
most closely resembles the $dR/dE_R \sim 1/E_R^2$  spectrum of hidden sector dark matter
(in the nuclear recoil energy region 2.5 keV$_{NR}$ - 5.0 keV$_{NR}$).  
}
\end{table}

\section{Germanium spectrum - modulated part}

The dark matter interaction rate can be expanded in terms of an unmodulated and modulated part:
\begin{eqnarray}
{dR \over dE_R} = {dR_0 \over dE_R} + {dR_1 \over dE_R} \ \cos \omega (t - t_0)
\end{eqnarray}
where $\omega = 2\pi/T$, $T = 1$ year and $t_0 = 152.5$ days.
The previous section has dealt with the unmodulated part.
Here we examine the expectations for the annual modulation amplitude, $dR_1/dE_R$.
We study first the simplest hidden sector case, considered in the previous section, where only the $F_2$ component
is heavy enough to give significant contributions.
In figure 5 we give the predicted annual modulation amplitude, $dR_1/dE_R$,
for various values of $m_{F_2}$.
Also shown is the annual modulation amplitude  
for WIMP dark matter with $m_\chi = 8.5$ GeV. 
The cross-sections are normalized as in figures 1-3.

CoGeNT has reported some positive hints for an annual modulation
in their event rate\cite{cogent}. Meanwhile, CDMS\cite{cdmsmod} has constrained
any modulation above $5\ {\rm keV}_{NR}$, to be less than around $0.05$ ${\rm counts/kev}_{NR}/{\rm kg}/{\rm day}$.
Figure 5 shows that all of the 
models considered predict a very small modulation amplitude above $5\ {\rm keV}_{NR}$, well within CDMS constraints.
Figure 5 also indicates that the (minimal) hidden sector dark matter 
model considered predicts an annual modulation of negative sign at the lowest energies. This is not
supported by CoGeNT's initial measurements\cite{cogent}, which hint at a sizable positive amplitude 
$dR_1/dE_R \sim 0.2$ counts/keV$_{NR}$/kg/day
averaged over the energy range $2.5 \ {\rm keV}_{NR} < E_R < 5 \ {\rm keV}_{NR}$.

\vskip 0.1cm
\centerline{\epsfig{file=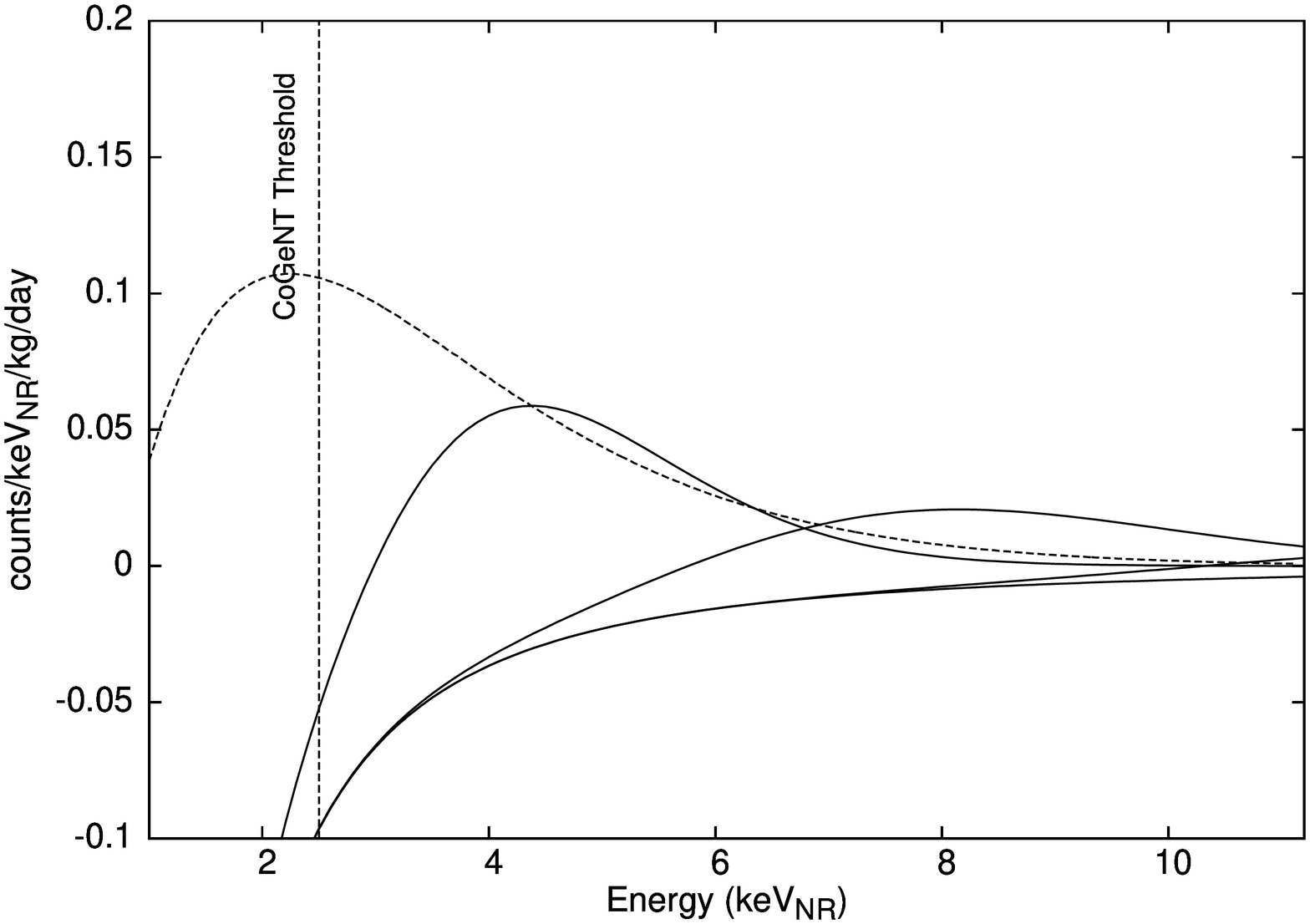,angle=270,width=12.9cm}}
\vskip 0.3cm
\noindent
{\small
Figure 5: Annual modulation amplitude for $m_\chi = 8.5$ GeV WIMP (dashed-line) and hidden sector dark matter
with  $m_{F_2}/{\rm GeV}$ = 20, 30, 45 and 60 (solid lines
from top to bottom).  
[$v_{rot} = 230$ km/s and $\bar m = 1.0$ GeV].
}

\vskip 0.8cm

Hidden sector dark matter is expected to be multi-component and there might be light components giving positive 
contributions at low energies. 
Such a possibility is certainly expected in the specific mirror dark matter 
case. To
illustrate this possibility we therefore consider the  mirror dark matter model. 
Recall in that model a dark matter spectrum e$'$, H$'$, He$'$, O$'$, Fe$'$, ...
with masses identical to their ordinary matter counterparts is predicted.
The H$'$ and He$'$ components are too light to give an observable
signal in any of the existing experiments. This leaves only the heavier mirror `metal' components,
which it happens, have roughly the right masses to explain the data.

It is, of course,
very difficult to predict the heavy mirror element abundances in the Universe.
However, since mirror `metal' components are expected to be forged in mirror stars,
a logical starting point might be to examine the ordinary metal abundances in the Universe.
According to Wolfram\cite{alpha},
the eight most abundant metals in the Universe are: O, C, Ne, Fe, N, Si, Mg and S.
Their abundances are given in table 2.
If we assume that mirror metal abundances have a similar pattern then this
motivates considering a mirror particle spectrum dominated by just four elements:
\footnote{We arrive at these four elements by observing that
(a) the carbon contribution is suppressed via kinematic effects relative to oxygen 
and can be approximately discarded,
(b) {N,O} have similar mass which we approximate with O and
(c) {Mg, Si, S} have mass number $28\pm 4$, so that these can be roughly approximated by just Si. 
}
O$'$, Ne$'$, Si$'$, Fe$'$,
with metal mass fractions given by  
\begin{eqnarray}
(a) \ \ {\xi_{O'} \over {\cal N}}  = 0.7, \ {\xi_{Ne'} \over {\cal N}} = 0.1, \ 
{\xi_{Si'} \over {\cal N}} = 0.1, \ {\xi_{Fe'} \over {\cal N}} = 0.1 
\end{eqnarray}
where ${\cal N} \equiv \sum \xi_{A'}$ and the sum runs over all mirror elements except H$'$, He$'$.
We also consider two alternative
examples. The first gives more weight to  heavier components, while the second
more to lighter components:
\begin{eqnarray}
&(b)& \ \  {\xi_{O'} \over {\cal N}}  = 0.5, \ {\xi_{Ne'} \over {\cal N}} = 0.05, \ 
{\xi_{Si'} \over {\cal N}} = 0.2, \ {\xi_{Fe'} \over {\cal N}} = 0.25\  , \nonumber \\ 
&(c)& \ \  {\xi_{O'} \over {\cal N}}  = 0.8, \ {\xi_{Ne'} \over {\cal N}} = 0.12, \ 
{\xi_{Si'} \over {\cal N}} = 0.08, \ {\xi_{Fe'} \over {\cal N}} = 0\  . \nonumber \\ 
\end{eqnarray}
It is expected that
more stellar processing in larger mirror stars should increase the proportion of heavier metal
components. The three sets (a), (b) and (c) (above) aim to illustrate the range of possible
values for the abundances. More extreme ranges are, of course, possible.

\begin{table}
\centering
\begin{tabular}{c c}
\hline\hline
element  & metal mass fraction  \\
\hline
Oxygen [O]  & 0.48   \\
Carbon [C]  & 0.24   \\
Neon [Ne]   & 0.07   \\
Iron [Fe] & 0.06 \\
Nitrogen [N] & 0.05 \\
Silicon [Si] & 0.04 \\
Magnesium [Mg] & 0.03 \\
Sulfur [S] & 0.02 \\
\hline\hline
\end{tabular}
\caption{
The eight most abundant metals in the Universe. The metal mass fraction of nuclei $A_1$ is $\xi_{A_1}/\sum \xi_{A}$ where the sum
runs over all elements except H and He. }
\end{table}

In figure 6a,b we plot the Germanium recoil spectrum and annual modulation amplitude for mirror dark matter with the
three different sets of abundances $(a), \ (b)$ and $(c)$.
We assume $\bar m = 1$ GeV and 
$v_{rot} = 230$ km/s for the (a) abundance set,
$v_{rot} = 200$ km/s for the (b) abundance set and
$v_{rot} = 270$ km/s for the (c) abundance set. 
We also show the corresponding spectra for $m_\chi = 8.5$ GeV WIMPs ($v_{rot} = 230$ km/s).

\vskip 0.4cm
\centerline{\epsfig{file=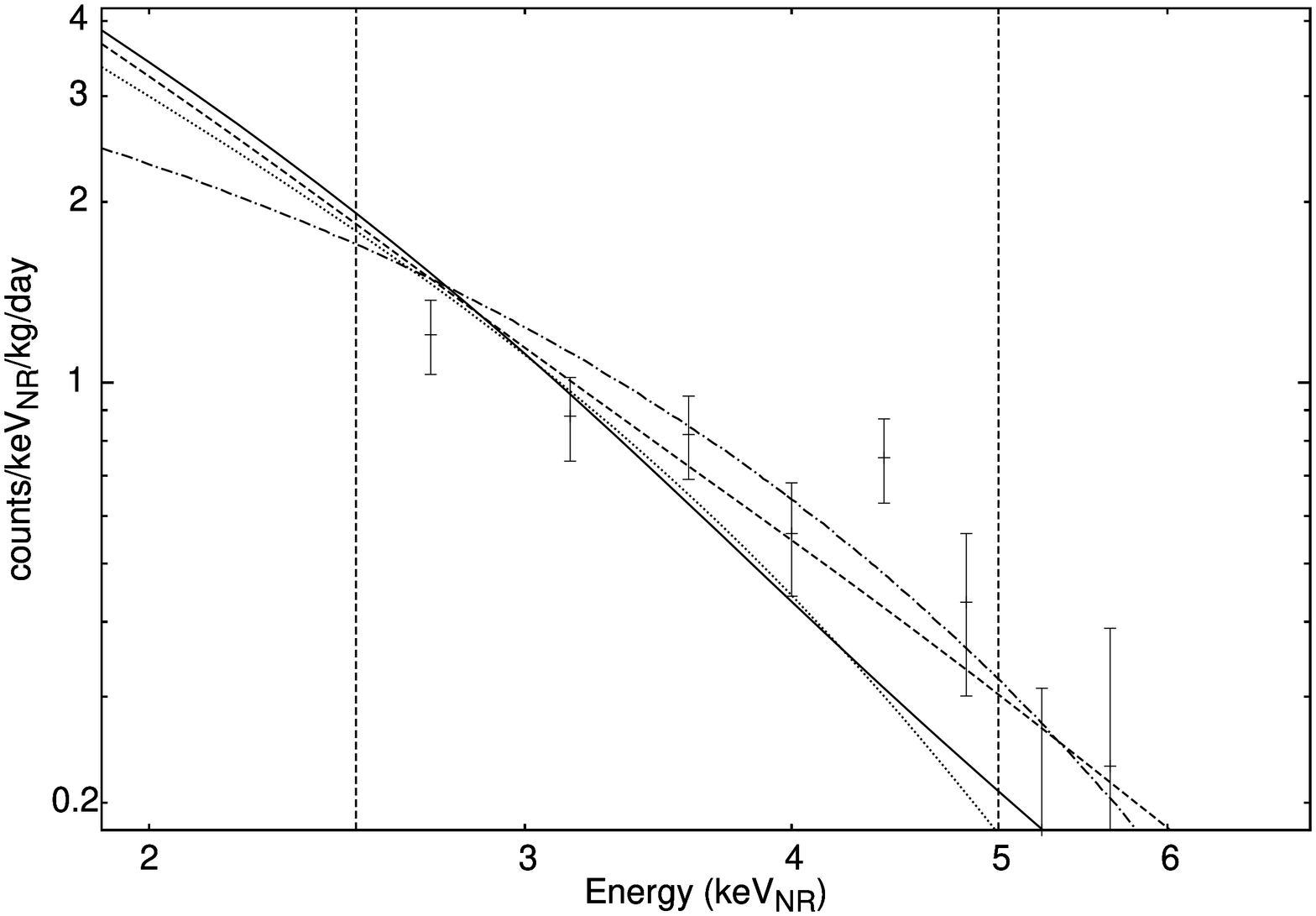,angle=270,width=12.9cm}}
\vskip 0.2cm
\noindent
{\small
Figure 6a: Germanium spectrum with mirror dark matter
for: the $(a)$ abundances [solid line], $(b)$ abundances [dashed line]  and
$(c)$ abundances [dotted line] (see text).
Also shown is the spectra for $m_\chi = 8.5$ GeV WIMP model [dashed-dotted line].
The parameters $\epsilon$ and $\sigma_n$ are adjusted so that the curves have similar
normalization in the CoGeNT signal region.
}
\vskip 0.9cm
\centerline{\epsfig{file=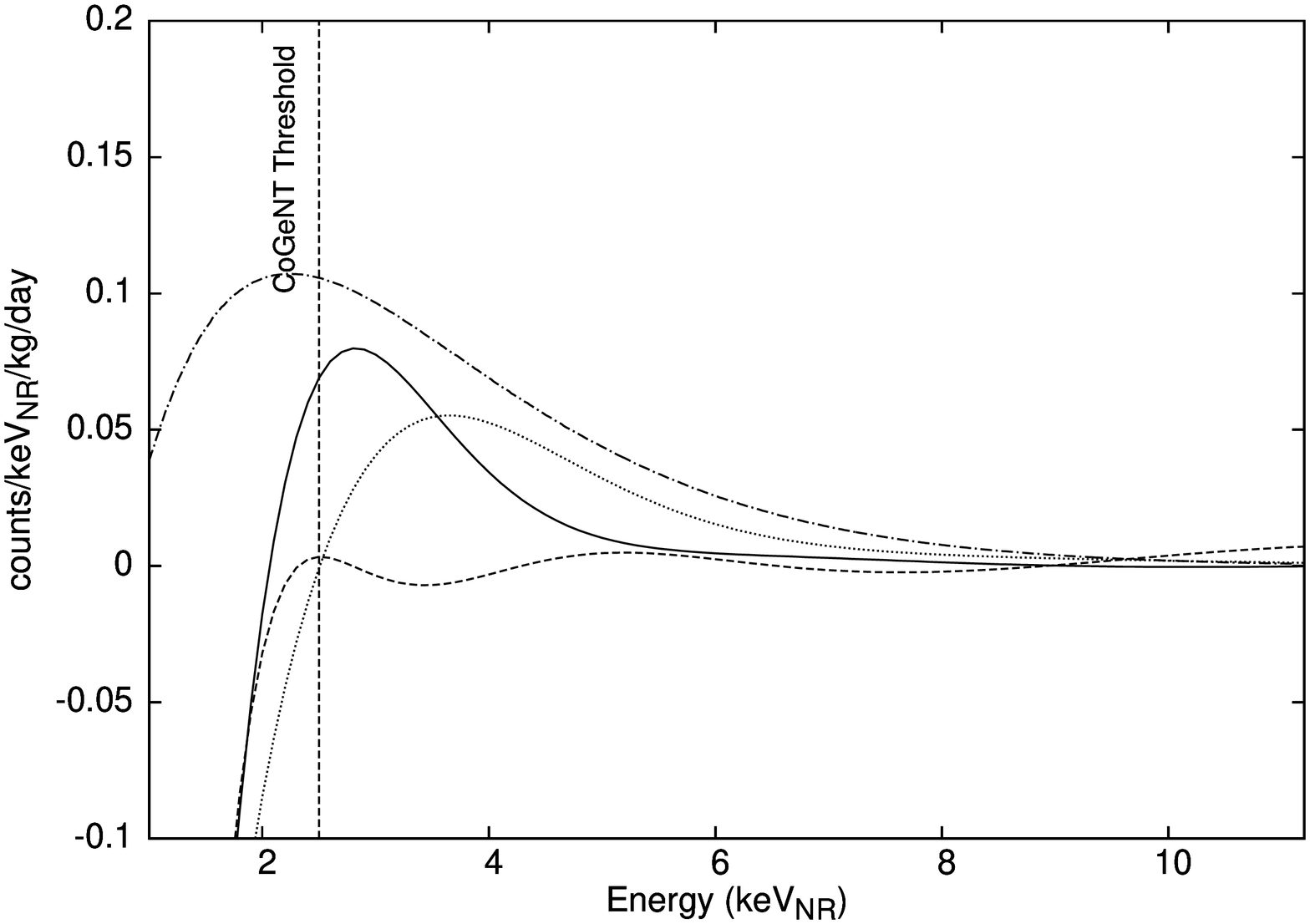,angle=270,width=12.9cm}}
\vskip 0.2cm
\noindent
{\small
Figure 6b: Annual modulation amplitude for the same cases and parameters as figure 6a.
}

\vskip 0.9cm

Figures 6a,b indicate that each of these examples can fit the data, with the $(b)$ abundances slightly preferred
by CoGeNT's spectrum. In each case the annual modulation amplitude is quite small in CoGeNT, with
the $(a), \ (c)$ abundances able to give a small positive contribution at low energies [essentially this is due to the
larger O$'$ proportion].
Each of these examples has just
one free parameter, which can be taken as the product: $\epsilon \sqrt{\xi_{O'}}$. 
The normalization of the CoGeNT spectrum 
fixes this free parameter:  $\epsilon \sqrt{\xi_{O'}} \approx 4 \times 10^{-10}$ for the $(a), \ (c)$
abundance set and $\epsilon \sqrt{\xi_{O'}} \approx 2.4 \times 10^{-10}$ for the $(b)$ abundance set.
With this parameter fixed, I have checked that the examples with $(a)$ and $(b)$ abundances also give a reasonable fit 
to the DAMA annual modulation spectrum 
and the CRESST-II excess (when analysed as in ref.\cite{foot2012a}).
The example with $(c)$ abundance can fit the DAMA annual modulation signal but gives too few  
events for CRESST-II.
These examples illustrate the substantial parameter space whereby mirror dark matter 
can simultaneously explain the three positive dark matter signals. 

Although the size of the annual modulation signal predicted in figures 5 and 6b is fairly small, there are nevertheless excellent prospects that
it can be measured (or strongly constrained).
The C-4\cite{c4}, 
CDEX\cite{cdex}, and the MAJORANA demonstrator\cite{maj}
experiments
plan Germanium 
target mass $\sim $ 4 kg, 10 kg and 40 kg
respectively. This is orders of magnitude larger than CoGeNT's $\sim 0.3$ kg target.
This should be sufficient to distinguish hidden sector dark matter from light WIMPs via
the annual modulation signal. This should complement the information obtained from precise measurements of the unmodulated 
recoil spectrum considered in the previous section.

\section{Conclusion}

Current data from the DAMA, CoGeNT and CRESST-II experiments can potentially be 
explained within several dark matter frameworks.  
The most promising of which appears to be 
light WIMP dark matter and hidden sector dark matter.
Both of these approaches feature spin independent elastic scattering of
dark matter particles on nuclei.
However light WIMP 
dark matter invokes a single particle species that interacts with ordinary matter via contact interactions,
while hidden sector dark matter is typically 
multi-component and is assumed to interact via the exchange of a massless mediator.

We have examined and compared the predictions of hidden sector dark matter with that of light WIMPs  
for experiments using Germanium detectors, such as CoGeNT, and 
in the near future, C-4, CDEX and the MAJORANA demonstrator.
Hidden sector dark matter predicts a spectrum: $dR/dE_R \propto 1/E_R^2$ 
or even more steeply falling if the threshold of light components is passed.
Scattering of
light WIMPs, on the other hand, produce a recoil
spectrum which depends sensitively on the WIMP mass $m_\chi$, with $dR/dE_R \to $ constant as $E_R \to 0$.
For the energy range of CoGeNT's excess, 
WIMPs with $m_\chi \simeq 8.5$ GeV produce a recoil spectrum that approximately matches the 
$\sim 1/E_R^2$\ \ dependence characteristic of hidden sector models.
The WIMP model fit to the CoGeNT excess, and also an analysis\cite{cf} of low energy CDMS
data, favour $m_\chi \approx 8.5$ GeV. This `coincidence' can be viewed as 
interesting evidence in support of hidden sector dark matter.

Future experiments should be able to differentiate  
hidden sector dark matter from light WIMPs, even for a WIMP mass of 8.5 GeV.
The C-4 experiment\cite{c4} for instance, which aims to have a lower threshold and reduced
background should be able to probe the signal over a significantly wider
energy range. The characteristic flattening of the predicted WIMP
spectrum at low energies, absent for hidden sector
dark matter, is one way. Another way is via
the annual modulation signal. We have shown that the two scenarios 
generally give different annual modulation spectra. This should provide another means
to distinguish light WIMPs from hidden sector dark matter.

\vskip 0.7cm
\noindent
{\large \bf Acknowledgments}

\vskip 0.1cm
\noindent
The author thanks M. Schmidt for a useful comment.
This work was supported by the Australian Research Council.


\begin{thebibliography}{99}

\bibitem{dama1}
R.~Bernabei {\it et al.},
Riv.\ Nuovo Cim.\  {\bf 26N1}, 1 (2003)  [astro-ph/0307403];  
Int.\ J.\ Mod.\ Phys.\ D {\bf 13}, 2127
(2004)  [astro-ph/0501412];  
Phys.\ Lett.\ B {\bf 480}, 23 (2000).  

\bibitem{dama2}
R.~Bernabei {\it et al.}  [DAMA and LIBRA Collaborations],
Eur.\ Phys.\ J.\ C {\bf 67}, 39 (2010)  [arXiv:1002.1028];  
Eur.\ Phys.\ J.\ C {\bf 56}, 333 (2008)  [arXiv:0804.2741].

\bibitem{cogent}
C.~E.~Aalseth {\it et al.} [CoGeNT Collaboration],
Phys.\ Rev.\ Lett.\  {\bf 107}, 141301 (2011)  [arXiv:1106.0650];  
Phys.\ Rev.\ Lett.\  {\bf 106}, 131301 (2011)  [arXiv:1002.4703].  


\bibitem{cogent2}
C.~E.~Aalseth {\it et al.}  [CoGeNT Collaboration],
arXiv:1208.5737.

\bibitem{cresst-II}
G.~Angloher {\it et al.},
Eur.\ Phys.\ J.\ C {\bf 72}, 1971 (2012) [arXiv:1109.0702].

\bibitem{dm}
A.~K.~Drukier, K.~Freese and D.~N.~Spergel,
Phys.\ Rev.\ D {\bf 33}, 3495 (1986).
K.~Freese, J.~A.~Frieman and A.~Gould,
Phys.\ Rev.\ D {\bf 37}, 3388 (1988).



\bibitem{c4}
R.~M.~Bonicalzi {\it et al.}  [C-4 Collaboration],
arXiv:1210.6282.

\bibitem{cdex}
Q.~Yue {\it et al.}  [CDEX-TEXONO Collaboration],
arXiv:1201.5373.

\bibitem{maj}
G.~K.~Giovanetti {\it et al.},
J.\ Phys.\ Conf.\ Ser.\  {\bf 375}, 012014 (2012).

\bibitem{review}
J.~D.~Lewin and P.~F.~Smith,
Astropart.\ Phys.\  {\bf 6}, 87 (1996).

\bibitem{c1}
D.~Hooper, J.~I.~Collar, J.~Hall, D.~McKinsey and C.~Kelso,
Phys.\ Rev.\ D {\bf 82}, 123509 (2010)
[arXiv:1007.1005].

\bibitem{gondolo}
C.~Savage, G.~Gelmini, P.~Gondolo and K.~Freese,
JCAP {\bf 0904}, 010 (2009) [arXiv:0808.3607];
P.~Gondolo and G.~Gelmini,
Phys.\ Rev.\ D {\bf 71}, 123520 (2005) [hep-ph/0504010].

\bibitem{wimp}
C.~Savage, G.~Gelmini, P.~Gondolo and K.~Freese,
Phys.\ Rev.\ D {\bf 83}, 055002 (2011) [arXiv:1006.0972];
Y.~Mambrini,
JCAP {\bf 1009}, 022 (2010) [arXiv:1006.3318];
P.~J.~Fox, J.~Liu and N.~Weiner,
Phys.\ Rev.\ D {\bf 83} (2011) 103514
[arXiv:1011.1915];
J.~L.~Feng, J.~Kumar, D.~Marfatia and D.~Sanford,
Phys.\ Lett.\ B {\bf 703}, 124 (2011)
[arXiv:1102.4331];
C.~Arina, J.~Hamann and Y.~Y.~Y.~Wong,
JCAP {\bf 1109}, 022 (2011) [arXiv:1105.5121];
D.~Hooper and C.~Kelso,
Phys.\ Rev.\ D {\bf 84}, 083001 (2011) [arXiv:1106.1066];
P.~Belli, R.~Bernabei, A.~Bottino, F.~Cappella, R.~Cerulli, N.~Fornengo and S.~Scopel,
Phys.\ Rev.\  D {\bf 84}, 055014 (2011) [arXiv:1106.4667];
T.~Schwetz and J.~Zupan,
JCAP {\bf 1108}, 008 (2011) [arXiv:1106.624];
M.~Farina, D.~Pappadopulo, A.~Strumia and T.~Volansky,
JCAP {\bf 1111}, 010 (2011)
[arXiv:1107.0715];
J.~Kopp, T.~Schwetz and J.~Zupan,
JCAP {\bf 1203}, 001 (2012) [arXiv:1110.2721];
C.~Kelso, D.~Hooper and M.~R.~Buckley,
Phys.\ Rev.\ D {\bf 85}, 043515 (2012) [arXiv:1110.5338];
M.~T.~Frandsen, F.~Kahlhoefer, C.~McCabe, S.~Sarkar and K.~Schmidt-Hoberg,
JCAP {\bf 1201}, 024 (2012) [arXiv:1111.0292];
J.~M.~Cline, Z.~Liu and W.~Xue,
arXiv:1207.3039 [hep-ph].  


\bibitem{foot69}
R.~Foot,
Phys.\ Rev.\ D {\bf 69}, 036001 (2004) [hep-ph/0308254];  
Mod.\ Phys.\ Lett.\ A {\bf 19}, 1841 (2004)  [astro-ph/0405362];  
astro-ph/0403043;
Phys.\ Rev.\ D {\bf 82}, 095001 (2010) [arXiv:1008.0685];  
Phys.\ Lett.\ B {\bf 692}, 65 (2010)  [arXiv:1004.1424];  
Phys.\ Lett.\ B {\bf 703}, 7 (2011)  [arXiv:1106.2688].

\bibitem{he}
R.~Foot and X-G. ~He, Phys. \ Lett. \ B{\bf 267}, 509 (1991). 


\bibitem{flv}
R. Foot, H. Lew and R. R. Volkas, Phys. \ Lett. \ B{\bf 272}, 67 (1991);
Mod. \ Phys. \ Lett. \ A{\bf 7}, 2567 (1992);
R.~Foot and R.~R.~Volkas,
Phys.\ Rev.\ D {\bf 52}, 6595 (1995) [hep-ph/9505359].

\bibitem{foot2012a}
R.~Foot,
Phys.\ Rev.\ D {\bf 86}, 023524 (2012).

\bibitem{foot2012b}
R.~Foot,
arXiv:1209.5602.

\bibitem{cf}
J.~I.~Collar and N.~E.~Fields,
arXiv:1204.3559.

\bibitem{holdom} 
B.~Holdom,
Phys.\ Lett.\ B {\bf 166}, 196 (1986).

\bibitem{ital}
N.~Fornengo, P.~Panci and M.~Regis,
Phys.\ Rev.\ D {\bf 84}, 115002 (2011)
[arXiv:1108.4661].

\bibitem{feng} 
J.~L.~Feng, M.~Kaplinghat, H.~Tu and H.~-B.~Yu,
JCAP {\bf 0907}, 004 (2009) [arXiv:0905.3039].

\bibitem{kali}
N.\ F. \ Bell, K. \ Petraki, I.~M.~Shoemaker and R.~R.~Volkas,
Phys.\ Rev.\ D {\bf 84}, 123505 (2011)  [arXiv:1105.3730];
K.~Petraki, M.~Trodden and R.~R.~Volkas,
JCAP {\bf 1202}, 044 (2012) [arXiv:1111.4786].  

\bibitem{cline}
J.~M.~Cline, Z.~Liu and W.~Xue,
Phys.\ Rev.\ D {\bf 85}, 101302 (2012) [arXiv:1201.4858].

\bibitem{sph}
R.~Foot and R.~R.~Volkas,
Phys.\ Rev.\ D {\bf 70}, 123508 (2004)  [astro-ph/0407522].  

\bibitem{xenon100}
E.~Aprile {\it et al.}  [XENON100 Collaboration],
arXiv:1207.5988.

\bibitem{cdms}
Z.~Ahmed {\it et al.}  [CDMS-II Collaboration],
Science {\bf 327}, 1619 (2010) [arXiv:0912.3592].

\bibitem{review2}
A.~Y.~Ignatiev and R.~R.~Volkas,
hep-ph/0306120;
R.~Foot,
Int.\ J.\ Mod.\ Phys.\ D {\bf 13}, 2161 (2004)  [astro-ph/0407623];  
Int.\ J.\ Mod.\ Phys.\ A {\bf 19}, 3807 (2004) [astro-ph/0309330];  
Z.~Berezhiani,
Int.\ J.\ Mod.\ Phys.\ A {\bf 19}, 3775 (2004)  [hep-ph/0312335];  
P.~Ciarcelluti,
Int.\ J.\ Mod.\ Phys.\ D {\bf 19}, 2151 (2010)  [arXiv:1102.5530].  

\bibitem{paolo2}
P.~Ciarcelluti and R.~Foot,
Phys.\ Lett.\ B {\bf 690}, 462 (2010)  [arXiv:1003.0880].  




\bibitem{bmei}
D.~Barker and D.~-M.~Mei,
arXiv:1203.4620.

\bibitem{cdmslow}
Z.~Ahmed {\it et al.}  [CDMS-II Collaboration],
Phys.\ Rev.\ Lett.\  {\bf 106}, 131302 (2011) [arXiv:1011.2482].

\bibitem{cdmsmod} 
Z.~Ahmed {\it et al.}  [CDMS Collaboration],
arXiv:1203.1309. 

\bibitem{alpha}
http://www.wolframalpha.com/

\end{thebibliography}
\end{document}